%% file: SPAWC_ISAC.tex
\documentclass[conference]{IEEEtran}

\input{input.tex}
\usepackage{epstopdf}
\usepackage{enumerate}
\usepackage{algorithm}
\usepackage{amsmath}
\usepackage{float}
\usepackage{color}
\usepackage{makeidx}
\usepackage{bm}
\usepackage[hidelinks=True]{hyperref}
\usepackage{cleveref}
\usepackage{url}
\usepackage{steinmetz}
\usepackage{varwidth}
\usepackage{soul}
\graphicspath{{/}}
\usepackage{balance}

\newcommand{\normsq}[1]{\left\lVert#1\right\rVert^2}

\newcommand{\absq}[1]{\left\lvert#1\right\rvert^2}

\begin{document}
	
\title{Learning Beamforming in Cell-Free \\ Massive MIMO ISAC Systems}
	
\author{Umut Demirhan and Ahmed Alkhateeb%
}
	\maketitle
	
\renewcommand{\thefootnote}{}
\footnotetext{The authors are with the School of Electrical, Computer and Energy Engineering, Arizona State University, Tempe, AZ, 85281 USA (Email: udemirhan, alkhateeb@asu.edu).}

\renewcommand{\thefootnote}{\arabic{footnote}}

	\begin{abstract}
            Beamforming design is critical for the efficient operation of  integrated sensing and communication (ISAC) MIMO systems. ISAC beamforming design in cell-free massive MIMO systems, compared to colocated MIMO systems, is more challenging due to the additional complexity of the distributed large number of access points (APs). To address this problem, this paper first shows that graph neural networks  (GNNs) are a suitable machine learning framework. Then, it  develops a novel heterogeneous GNN model inspired by the specific characteristics of the cell-free ISAC MIMO systems. This model enables the low-complexity scaling of the cell-free ISAC system and does not require full retraining when additional APs are added or removed. Our results show that the proposed architecture can achieve near-optimal performance, and applies well to various network structures.
	\end{abstract}

	\section{Introduction}

    Integrated sensing and communications (ISAC) is envisioned as an integral part of next-generation and future communication systems by integrating sensing into the communication systems \cite{liu2022integrated}. With this integration, the communication systems can sense the environment and provide information that can be utilized for various purposes. These purposes include improving communication service quality through beamforming and blockage prediction, providing up-to-date traffic information, and healthcare and security applications. Despite these advantages, the ever-increasing density of the communication networks makes integrating sensing challenging due to the interference and increased overall complexity, which may be overcome with machine learning (ML) \cite{Demirhan_mgazine_radar}. This, however, is not a trivial task, and ML models tailored for specific problems and adaptable to various networks are needed. With this motivation, we investigate machine learning for beamforming in cell-free massive MIMO ISAC systems.

    Cell-free massive MIMO \cite{ngo2017cell} is a critical paradigm for future wireless communication networks, where the large interference due to the densification of the networks is managed with the coordination between the access points (APs). Specifically, in cell-free massive MIMO, a large number of APs serve the user equipments (UEs), jointly with coordination between them. This coordination further allows taking advantage of the variety in the channels between different APs and a UE for a seamless performance. In the original work \cite{ngo2017cell}, the advantage of cell-free massive MIMO was shown against the conventional cell-based networks. After that, significant research has been carried out for the various properties of cell-free massive MIMO, such as scaling, precoding, and fronthaul limitation \cite{demirhan2022enabling}. This interest has further extended to the integration of sensing into communications, and power allocation \cite{behdad2022power} and beamforming \cite{demirhan2023cell} in cell-free ISAC has been studied. These problems, however, mainly relied on the solutions to the complex optimization problems, which are difficult to apply in real time and hard to scale to various networks. To that end, in this paper, we aim to propose a scalable beamforming solution for cell-free ISAC networks.

    Graph neural networks (GNNs) are specialized neural networks that take advantage of the dependence of graphs and can generalize graph-based problems \cite{zhou2020graph}. Specifically, GNNs can efficiently learn the graph-based models and effortlessly extend the learned models to various graphs by scaling over the nodes and edges. Therefore, GNNs provide an essential tool for the problems in wireless communications, where the wireless networks constitute the graphs for the GNNs. The critical but challenging problems in wireless networks, such as power and bandwidth allocation, interference management, and distributed beamforming, can benefit from utilizing GNNs with their potential to adapt to various networks.
    
    With this motivation, in this paper, we aim to tailor a GNN solution for the ISAC beamforming problem in cell-free massive MIMO networks. We first define our ISAC beamforming objective, which aims to maximize the sum of the user communication rates and log of the sensing SNR. We then design a heterogeneous edge graph neural network model to tackle the proposed JSC beamforming optimization problem. In this model, tailored for the cell-free ISAC systems, we adopt three types of nodes: AP, UE, and sensing target. The edges connect the APs to UEs and the sensing target, respectively, representing the communication and sensing channels between these nodes. This design further allows utilizing the same coefficients in various network configurations, enabling the scaling for cell-free systems. Our evaluations show that the proposed network model can achieve the optimal performance for only communication, and performs well to provide a trade-off between the communication and sensing functions. It further provides a reasonable scaling capability for the APs without retraining the model, allowing adaptability to the changes in the network.

	\section{System Model} \label{sec:systemmodel}

\begin{figure}[t]
		\centering
		\includegraphics[width=1\linewidth]{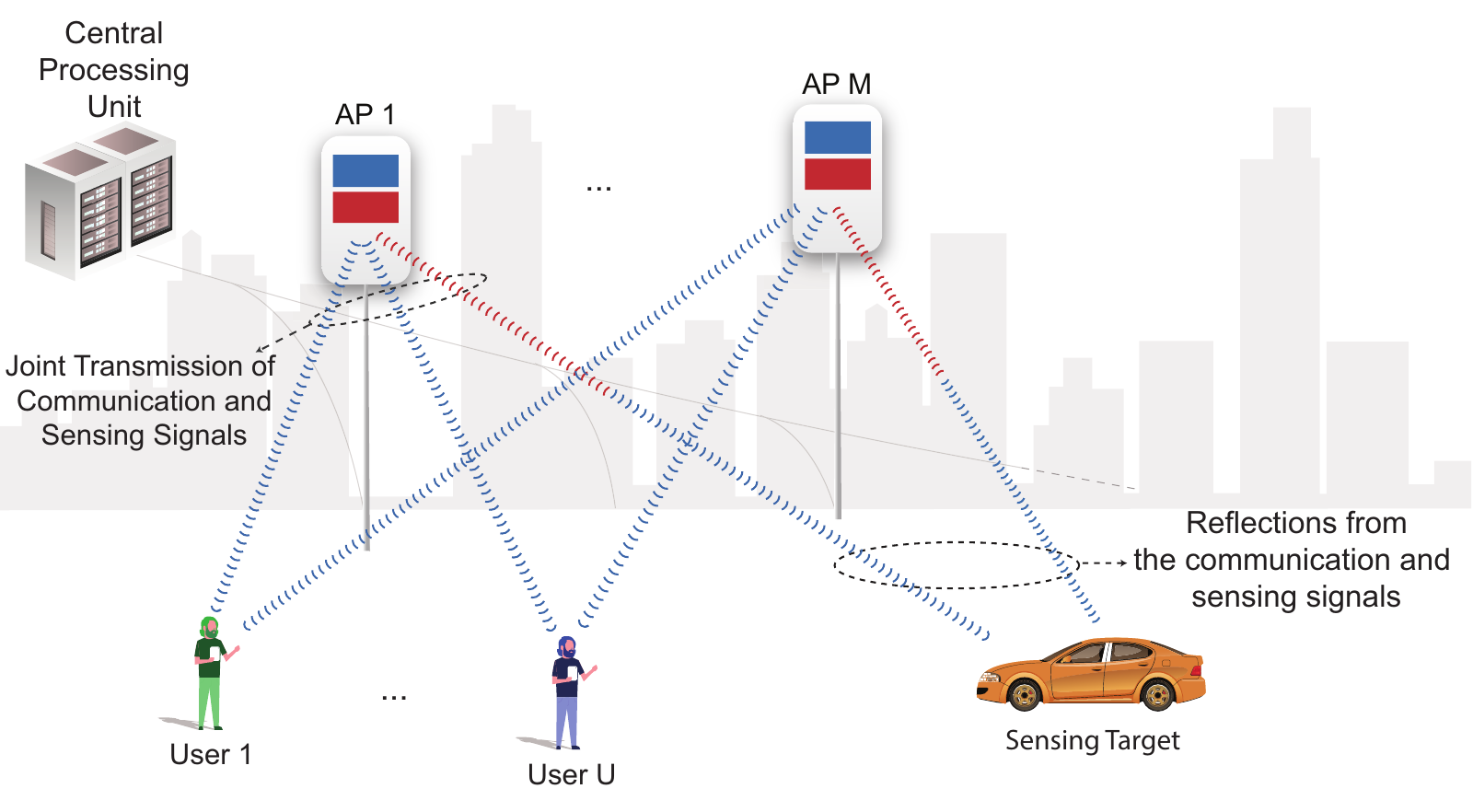}
		\caption{The APs transmit signals with beamforming to jointly serve the UEs and sense the target. The signals reflected from the target are collected back at the APs for multistatic sensing.}
		\label{fig:systemmodel}
	\end{figure}

	We consider a cell-free massive MIMO ISAC system with $M$ access points (APs) and $U$ communication users, as illustrated in \figref{fig:systemmodel}. In the downlink, the set of APs $\mathcal{M}$ transmit communication and sensing waveforms to serve the $U$ users jointly, and to sense the environment. Specifically, these APs, $\mathcal{M}$, also receive the possible reflections/scattering of the transmitted waveforms on the various targets/objects in the environment. Note that the APs are simultaneously transmitting and receiving signals. The APs are equipped with $N_t$ transmitter antennas to serve the UEs, and $N_r$ receiver antennas to sense the environment. In addition, we assume that all the APs have digital beamforming capabilities, i.e., each antenna element has a dedicated radio frequency (RF) chain. The UEs are equipped with single antennas. The APs are connected to a central unit that allows joint design and processing, and they are assumed to be fully synchronized. 
	
	\subsection{Signal Model}
	In this subsection, we define the joint sensing and communication signal model for the downlink transmissions. The APs jointly transmit $U$  communication streams, $\{x_u[\ell]\}_{u\in\mathcal{U}}$, and $Q$ sensing streams, $\{x_q[\ell]\}_{q\in\mathcal{Q}}$, where $\mathcal{Q}=\{U+1, \ldots, U+Q\}$ and with $\ell$ denoting the $\ell$'s symbol in these communication/sensing streams. For ease of exposition, we also define the overall set of streams as $\mathcal{S}=\mathcal{U}\cup\mathcal{Q}=\{1, \ldots, S\}$ with $S=U+Q$. If $\bx_{m}[\ell] \in \mathbb{C}^{N_t \times 1}$ denotes the transmit signal from the transmitting AP $m$ due to the $\ell$-th symbol, we have
	\begin{equation} \label{eq:transmitsymbol}
		\bx_{m}[\ell] = \underbrace{\sum_{u \in \, \mathcal{U} }\bff_{m u} x_u[\ell]}_{\textrm{Communication}} + \underbrace{\sum_{q \in \mathcal{Q}} \bff_{m q} x_q[\ell]}_{\textrm{Sensing}} = \sum_{s \in \mathcal{S}} \bff_{m s} x_s[\ell],
	\end{equation}
	where $x_s[\ell] \in\mathbb{C}$ is the $\ell$-th symbol of the $s$-th stream, $\bff_{m s} \in \mathbb{C}^{N_t \times 1}$ is the beamforming vector for this stream applied by AP $m$. The symbols are assumed to be of unit average energy, $\bE[|x_s|^2]=1$. The beamforming vectors are subject to the total power constraint at each AP, $P_{m}$, given as
	\begin{equation}
		\bE[\normsq{\bx_{m}[\ell]}] = \sum_{s\in\mathcal{S}}\normsq{\bff_{m s}} \leq P_{m}.
	\end{equation}
	Further, by stacking the beamforming vectors of stream s of all the APs, we define the  beamforming vector $\bff_s = \begin{bmatrix} \bff^T_{1s} & \ldots & \bff_{M_t s}^T \end{bmatrix}^T \in \mathbb{C}^{M_t N_t}$.
	
	For each stream $s$, we denote the sequence of $L$ transmit symbols as $\bx_s = \begin{bmatrix} x_s[1], \ldots, x_s[L] \end{bmatrix}^T$. Given this notation, we assume that the radar and communication signals are statistically independent, i.e., $\bE[\bx_s \bx_s^H] = \bI$ and $\bE[\bx_s \bx_{s'}^H]=\bm{0}$ for $s, s' \in \mathcal{S}$ with $s \neq s'$ \cite{liu2020joint}.
	
	\subsection{Communication Model}
     We denote the communication channel between UE $u$ and AP $m$ as $\bh_{mu} \in \mathbb{C}^{N_t \times 1}$. Further, by stacking the channels between user $u$ and all the APs, we construct $\bh_{u} \in \mathbb{C}^{M_t N_t \times 1}$. We assume a block fading channel model, where the channel remains constant over the transmission of the $L$ symbols. With this, the communication rate of UE $u$ is written as
    \begin{equation}
        R_u = \log_2(1+\textrm{SINR}^\textrm{(c)}_u).
    \end{equation}
    where the SINR is given in \eqref{eq:SINR_u}, with $\sigma_u^2$ being the variance of the receiver noise \cite{demirhan2023cell}.
	\begin{figure*}[!tbh]
    \begin{equation} \label{eq:SINR_u}
        \displaystyle
        \textrm{SINR}^\textrm{(c)}_u = \frac{\absq{\sum_{m \in \mathcal{M}} \bh^H_{mu} \bff_{mu}}}{\sum_{u' \in \mathcal{U} \backslash \{u\}} \absq{\sum_{m \in \mathcal{M}} \bh^H_{mu}  \bff_{mu'}} +  \sum_{q \in \mathcal{Q}} \absq{ \sum_{m \in \mathcal{M}} \bh^H_{mu} \bff_{mq}} + \sigma_{u}^2}
    \end{equation}
    \end{figure*}

	\subsection{Sensing Model}
    For the sensing channel model, we consider a single-point reflector, as commonly adopted in the literature \cite{behdad2022power, demirhan2023cell}. Specifically, the transmit signal is scattered from the single-point reflector and received by the APs. With a single path model, the channel between the transmitting AP $m_t$ and the receiving AP $m_r$ through the reflector is defined as
	\begin{align}
		\begin{split}
			\bG_{m_t m_r} &= {\alpha}_{m_t m_r} \ba(\theta_{m_r}) \ba^H(\theta_{m_t}),
		\end{split}
	\end{align}
	where $\alpha_{m_t m_r} \sim \mathcal{CN}(0, \zeta_{m_t m_r}^2)$ is the combined sensing channel gain, which includes the effects due to the path-loss and radar cross section (RCS) of the target. We assume that $\alpha_{m_t m_r}$ is not known while its statistics are available, which can be achieved by targeting a particular location. The angles of departure/arrival of the transmitting AP $m_t$ and receiving AP $m_r$ from the point reflector are denoted by $\theta_{m_t}$ and  $\theta_{m_r}$. We consider the Swerling-I model for the sensing channel \cite{richards2010principles}, which assumes that the fluctuations of RCS are slow and the sensing channel does not change within the transmission of the $L$ sensing and communication symbols in $\bx_s$. With this model, the signal received at AP $m_r$ at instant $\ell$ can be written as
	\begin{align} \label{eq:radar_rx}
		\begin{split}
			\by^\textrm{(s)}_{m_r}[\ell] &= \sum_{m_t\in \mathcal{M}_t} \bG_{m_t m_r} \, \bx_{m_t}[\ell] + \bn_{m_r}[\ell]
		\end{split}
	\end{align}
	where $\bn_{m_r}[\ell]\in\mathbb{C}^{N_r}$ is the receiver noise at AP $m_r$ and has the distribution $\mathcal{CN}(0, \varsigma_{m_r}^2 \bI)$. With this model, we adopt the joint SNR of the received signals as the sensing objective. Note that the use of the joint SNR requires joint processing of the radar signal at the $\mathcal{M}$ sensing receivers. The sensing SNR can be written as
	\begin{align} \label{eq:SNRsensing}
		\begin{split}
			\textrm{SNR}^\textrm{(s)} &= \frac{ \sum_{m_r \in \mathcal{M}}  \sum_{m_t\in \mathcal{M}} \zeta_{m_t m_r} ^2  \sum_{s\in\mathcal{S}} \normsq{\ba^H(\theta_{m_t}) {\bff}_{ms}} }{ \sum_{m_r \in \mathcal{M}_r} \varsigma_{m_r}^2},
		\end{split}
	\end{align}
	where $\zeta_{m_t m_r}^2$ denotes the variance of the combined sensing channel gain and $\varsigma_{m_r}^2$ is the variance of the radar receiver noise. Note that the sensing SNR is scaled with the contribution of all the \textit{communication and sensing} streams. For the details of the derivation, please refer to \cite{demirhan2023cell}.

    \section{Problem Definition} \label{sec:problemdef}
    In the previous sections, the objectives for the sensing and communication functions are defined. Using these, we aim to optimize the beamforming coefficients to maximize the sensing and communication functions jointly. To that end, various problems could be formulated based on the application's requirements. In this paper, we focus on a sum of communication and sensing metrics. Specifically, for the communication, we adopt the sum rate as the objective. For the sensing, we adopt a logarithm-based expression, i.e., $\log_2(1+\textrm{SNR}^{\textrm{(s)}})$. This approach is inspired by mutual-information of the radar signals, which is similar to the communication data rate, that is derived using the mutual information between the transmit signal and received signal. It further balances the objective with the $\log$ fairness between the communication SINRs and sensing SNR. With these, we can write our main objective as
    \begin{subequations}
    \begin{align}
            \max_{\{\bff_{ms}\}} \quad & \sum_{u\in\mathcal{U}} R_u + \beta_s \log_2(1 + \mathrm{SNR}^\mathrm{(s)}) \label{eq:objective} \\
            \textrm{s.t.} \quad &\sum_{s \in \mathcal{S}} \normsq{\bff_{ms}} \leq P_m
    \end{align}
    \end{subequations}
    where the sensing weight parameter $\beta_s$ provides an adjustable balance between the communication and sensing utilities. For envisioned JSC systems, the primary objective is still to provide high data rates for communication with the secondary objective being the sensing. To that end, selecting a small value for $\beta_s$ to gain some sensing performance by minimally sacrificing the communication performance could be an interesting choice. Also, note that $\beta_s=0$ leads to the sum rate maximization for the cell-free communication systems. 
    
	\section{Heterogenous Graph Neural Networks for Cell-free ISAC MIMO Systems} \label{sec:GNN}

    To develop a learning approach to the problem presented in the previous section, we first note the graph structure of the cell-free MIMO systems, i.e., the network consists of APs and UEs with APs jointly serving UEs. This structure can be represented as a bipartite graph with UEs and APs as the nodes. This graph structure, however, cannot be directly generalized to the joint communication and sensing case. In particular, the APs jointly serve the UEs and also sense the target. Therefore, there are the UEs, APs, and also the sensing target (ST) as the nodes. In addition to the various types of nodes, it demands different edge types for the AP-UE and AP-ST connections. Such a structure also allows the exploitation of the sensing and communication objectives with different permutation invariance properties, where the APs and UEs can be permuted independently, making them amenable to GNNs.
        
    \subsection{Graph Definition}
\begin{figure}[t]
    		\centering
    		\includegraphics[width=.53\linewidth]{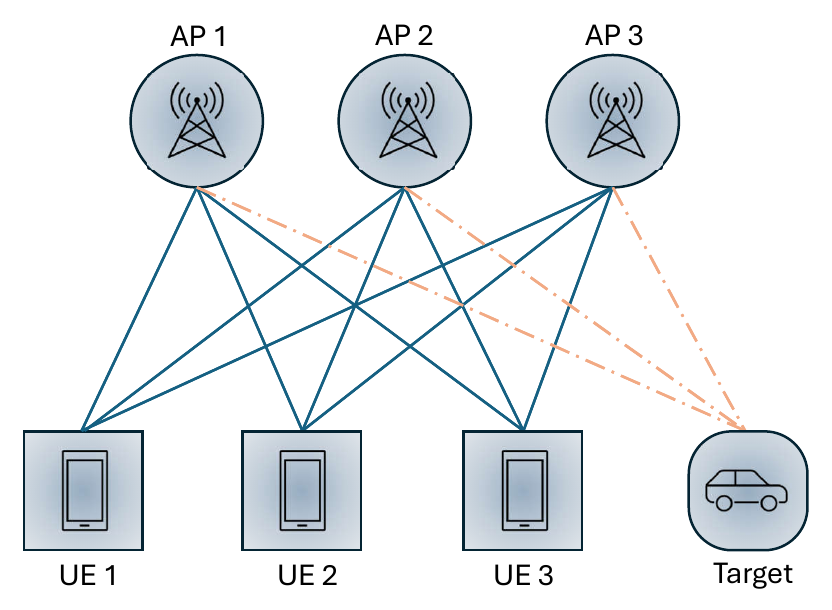}
    		\caption{The heterogeneous graph structure for the cell-free ISAC MIMO system is defined with three types of nodes (AP, UE, ST) and two types of edges (AP-UE and AP-ST).}
    		\label{fig:graphmodel}
    	\end{figure}

        A heterogeneous graph can be defined by a set of vertices and edges of various types, denoted by $\mathcal{G} = (\mathcal{V}, \mathcal{E}, \mathcal{V}_t, \mathcal{E}_t)$, where $\mathcal{V}$ and $\mathcal{E}$ are the sets of vertices and edges, while $\mathcal{V}_t$ and $\mathcal{E}_t$ denote their types. In cell-free ISAC MIMO systems, as shown in \figref{fig:graphmodel}, our vertices are defined with three types, i.e., $\mathcal{V}_t = \{\text{AP}, \text{UE}, \text{ST}\}$. For the edges, we define two types of bidirectional edges for the AP-UE and AP-ST links, given as $\mathcal{E}_t = \{(\text{AP-UE}), (\text{AP-ST})\}$. This structure allows the sensing and communication functions to be supported for learning: The sensing signals only add interference to communication, while the communication signals contribute to communication and sensing. The heterogeneous architecture enables the structure to support the joint objective. 

        \subsection{Graph Neural Networks}
        GNNs are the neural network models that can effectively exploit the underlying graph structure of the problem \cite{scarselli2008graph}. In GNNs, each vertex and edge is represented by a feature vector. Let us denote the feature for the vertex $i$ as $\bv_{i}$ and the feature for the edge between the vertices $i$ and $j$ as $\be_{ij}$. A GNN can be composed of multiple network layers. In each layer $l$, we denote the hidden representation with a superscript, i.e., $\bv_{i}^l$ and $\be^l_{ij}$. In a general way for the heterogenous GNNs, an update at the $l$-th layer of the GNN can be defined with two update equations: The edge update and the vertex update. We describe the details of these steps in the following.

        \noindent\textbf{Edge update:} At this step, each edge is updated based on the information of the same edge and connected vertices from the previous layer. If we define the update function for edge of edge type $t_e\in\mathcal{E}_t$ as $\psi^{t_e}(.)$, we can write
        \begin{equation}
            \be^l_{ij} = \psi^{t_e}(\bv^{l-1}_{i}, \be^{l-1}_{ij}, \bv^{l-1}_{j}).
        \end{equation}
        where the edge $(i,j)$ is of the type $t_e$. Note that with the heterogeneous structure, the edges of different types are updated with different update equations.
        
        \noindent\textbf{Vertex Update:} After the edges are updated, they are aggregated to be used for the vertex update. Specifically, the edges connected to the same vertex are first aggregated individually for each type of vertex. If we define the intermediate aggregation output for the $i$-th vertex (of type $t_v\in\mathcal{V}_t$) of the edges of type $t_e \in \mathcal{E}_t$ as $\ba^l_{i,t_e}$, we can write
        \begin{equation}
            \ba^l_{i, t_e} = \rho^{t_e, t_v}(\{\be^{l}_{ij}\}_{j \in \mathcal{N}(i),{j \in t_v},(i,j)\in t_e})
        \end{equation}
        where $\rho^{t_e, t_v}(.)$ is the aggregation function for the edges of type $t_e$ connected to the vertex of type $t_v$. $\mathcal{N}(i)$ denotes the neighbor (connected) vertices of vertex $i$. Then, the vertex can be updated by using the aggregated inputs of different edge types and vertex features in the previous layer. With a vertex update function $\phi^{t_v}(.)$ for a vertex of type $t_v$, we can write
        \begin{equation}
            \bv^l_i = \phi^{t_v}(\{\ba_{i,t_e}^l\}_{t_e \in \mathcal{E}_t}).
        \end{equation}

        \subsection{GNN Design for Cell-free ISAC MIMO Beamforming}
        As described in the previous subsection, it is possible to define different types of structures to obtain the beamforming coefficients. In our problem, however, it is interesting to utilize the edge embedding for the beamforming prediction. This allows scaling the edges with an increasing number of UEs or APs while the feature set at the AP, UE, and ST vertices/edges are of fixed size. In our design, we set the initial variables for each edge as the corresponding channel with
        \begin{equation}
            \be^0_{ij} = \begin{cases}
            \mathcal{S}(\bh^T_{mu}) & \text{if } (i,j) \in \{(\text{AP},\text{UE})\}\\
            \mathcal{S}(\ba(\theta_{m})) & \text{if } (i,j) \in \{(\text{AP},\text{ST})\}
            \end{cases}
        \end{equation}
        where $\mathcal{S}(\bx)=[\Re\{\bx^T\}, \Im\{\bx^T\}]^T$ is the stacking operation of real and imaginary parts. In addition to the standard update, we concatenate the initial value of each edge to the edge update at every layer. With the given initialization, this corresponds to expanding the edge features after the edge update $\be_{ij}^{l'}$ at layer $l$ as $\be^{l^T}_{ij} = [\be^{{l'}^T}_{ij} \  \be^{0^T}_{ij}]^T$. With this approach, each edge update also utilizes the channel coefficients (communication or sensing) in the features for the update. As the channel coefficients directly determine the performance for the communication and sensing, it is expected to improve the performance \cite{zhang2021scalable, kim2022bipartite}.
        
        With the initialization, we can update the network through the layers. We define the vertex update equations as
        \begin{equation}
            \bv^l_i = \sum_{t_e \in \mathcal{E}_t} \bW^{l-1}_{t_e,t_v} \cdot \textrm{concat}\big(\sum_{\substack{j \in \mathcal{N}(i)\\(i,j) \in t_e}}\be^{l-1}_{ij}, \bv^{l-1}_i\big)
        \end{equation}
        where the vertex feature is concatenated with the sum-aggregated edges features of a type. The weights $\bW^{l-1}_{t_e,t_v}$ denote the coefficients of the linear layer for the vertex of $t_v$ connected to the edge of type $t_e$. 
        At each edge, we apply the following equation with the corresponding weights
        \begin{equation}
            \be_{ij}^l = \textrm{activation}\bigg( \bW^l_{t_e} \cdot \textrm{concat}(\bv^l_i, \bv^l_j, \be^{l-1}_{ij}) \bigg)
        \end{equation}
        where $t_e$ is the type of the edge $(i,j)$, and $\bW^{l}_{t_e}$ is the weights of the corresponding linear layer. Note that there are only two types of edges, and hence, there are two weights at each layer.
        
        To satisfy the power constraints in our GNN model, we apply a normalization after each beamforming coefficient is obtained. In particular, after the output of the neural network is obtained without any constraints on the power, the weights corresponding to the beamforming coefficients at each AP are normalized to satisfy the power constraints. Mathematically, after the $L$-th (last) layer of the network, the beamforming coefficients are determined as $\hat{\bff}_{ms} = \sqrt{P_m} \, {\mathcal{S}^{-1}(\be^L_{ms})}({\sum_{s\in\mathcal{S}} \, \normsq{\be^L_{ms}} })^{-\frac{1}{2}}$, where $\mathcal{S}^{-1}$ converts the stacked real vector into complex vector as an inverse operation of $\mathcal{S}$. With this, the effect of the constraints is implicitly learned by the neural network since the cost function of the neural network is computed based on this normalized output.

        For the learning objective, it is important to adopt an approach requiring minimal training effort. To that end, we adopt an unsupervised learning approach, where the goal is to maximize objective function in \eqref{eq:objective} by learning weights.
        
	\section{Simulations} \label{sec:results}
	In this section, we evaluate the proposed solution's performance. Before the evaluation, we describe the simulation setup, ML parameters, and the baseline solution as follows. 
 
    \textbf{Simulation Setup:} The power at each AP is the same and the sum power of the APs is $1$, i.e., $P_m = \frac{1}{M}$. The noise power $\sigma^2=0.1$, providing a system-wide SNR of $10$dB. The communication channels are defined with a Rayleigh fading, i.e., $\bh_{mu} \sim \mathcal{CN}(0, \bI)$ and the sensing channels' variables are drawn from a uniform distribution, i.e., $\theta_{m} \sim \mathcal{U}[0, \frac{\pi}{2})$, with $\zeta_{m_t m_r} \sim \mathcal{CN}(0, 1)$. The number of antennas at each AP transmitter is taken as $N_t=8$ along with $U=2$ and $M=5$. 
 
    \textbf{Neural network parameters:} For the training and testing, we generate $100,000$ and $2,000$ samples for each parameter set. The networks are trained with the Adam algorithm using a learning rate of $4\cdot10^{-4}$ and batch size $256$. The models are trained for $60$ epochs and a cosine annealing scheduler. The activation function is selected as leaky ReLU with a negative slope $0.1$. The GNNs adopt $4$ hidden layers with $256$ dimensions each. Next, we describe the baseline methods.

    \textbf{Baseline methods:} In our evaluation, we adopt three different baseline approaches, namely, weighted minimum mean squared error (WMMSE) and conjugate beamforming (CB) for prioritizing communication, and CB for prioritizing the sensing performance: (i) \textit{WMMSE for communication:} In this baseline, we ignore the sensing objective by setting $\beta_s=0$ in the proposed problem and only aim at maximizing the communication sum rate. For the solution, we apply WMMSE as given in \cite{xu2023joint}. This solution provides an upperbound for communication performance. (ii) \textit{CB for communication:} As a simplified solution for the communication objective of the proposed problem, we apply conjugate beamforming at each AP with the powers of the beams normalized to satisfy the power constraints. In this case, the sensing beams are set as zero vectors with no power. (iii) \textit{CB for sensing}: To obtain a limit on the sensing performance, we apply conjugate of the target directions at each AP with the full power as the sensing beam, which provides an upperbound on the sensing.
    
\begin{figure}[t]
        \centering
        \includegraphics[width=.87\linewidth]{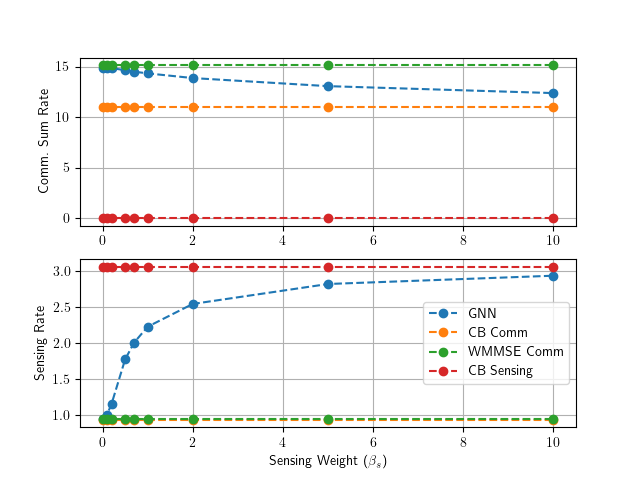}
        \caption{The performance of the neural networks trained with $U=2$ and $M=5$ for different sensing weight values.\vspace{-2ex}}
        \label{fig:sensingweight-U2M5}
    \end{figure}

    \textbf{Communication/Sensing trade-off:} We first evaluate the performance of the proposed GNN for various sensing weights, $\beta_s$, values. As can be seen from figure \figref{fig:sensingweight-U2M5} for $U=2$ and $M=5$, the proposed network converges well to the WMMSE communication for $\beta_s=0$, which shows the approximation capability of the communication part of the network. With increasing sensing weights, a trade-off is shown between the communication and sensing performance. It is important to note that with a small sacrifice in communication (e.g., $\beta_s=2$), the network achieves $85\%$ of the sensing of CB sensing solution, which allocates all the power to the sensing. It can further be seen that even with a large sensing weight ($\beta_s=10$), it is still possible to perform better than the CB comm solution while achieving almost the same performance as the CB sensing, which makes it interesting to adopt such a weighted JSC objective with the proposed solution.
 
\begin{figure}[t]
        \centering
        \includegraphics[width=.87\linewidth]{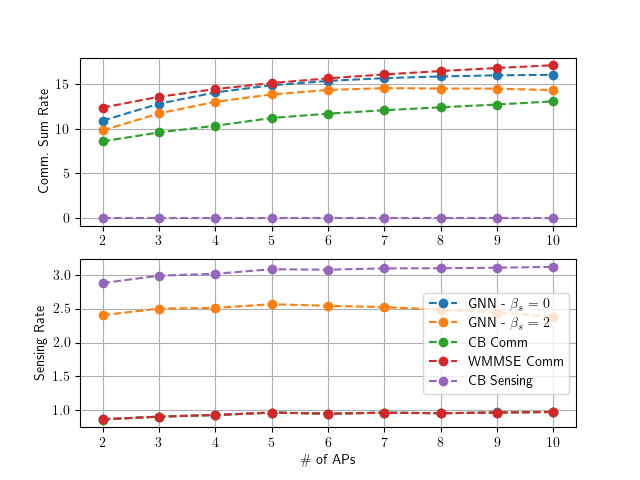}
        \caption{The proposed solution is trained with $U=2$ UEs and $M=5$ APs, and tested with different number of APs.\vspace{-2ex} 
        }
        \label{fig:Mscaling-U2M5}
    \end{figure}

    \textbf{Scaling with the number of UEs and APs:} An interesting property of the GNNs is the potential for adaptability of the model to different numbers of UEs and APs using the same parameters (without re-training the model). For the UEs, the solution only scales to the smaller number of UEs, while the performance of both sensing and communication degrades with the increasing number of UEs. To that end, it may be interesting to further tailor the GNN model for the UE scaling of the cell-free ISAC problem.
    For the scaling of APs, the resulting sensing and communication performance is shown \figref{fig:Mscaling-U2M5}. As observed from the figure, the network achieves the WMMSE communication performance for $M=5$, the number of APs it is trained for. For the communication-only scenario ($\beta_s=0$), with increasing and decreasing number of APs, the solution adapts well with a slight degradation with respect to the WMMSE communication solution. The JSC solution with $\beta_s=2$ shows a similar trend in the generalization. More interestingly, the sensing performance is sustained at almost a constant level for various numbers of APs, showing the capability of the proposed heterogeneous neural network for the sensing function.
    
    \section{Conclusion} \label{sec:conc}
        In this paper, we introduced ML for the challenging beamforming problem in cell-free massive MIMO ISAC systems. We specifically tailored a heterogeneous GNN architecture with the vertices as users, APs, and the sensing target. With this model, our results revealed that the proposed architecture can perform near-optimal and scale well for various number of APs, enabling ML-based beamforming for cell-free massive MIMO ISAC systems. Future research directions include exploration of the weight parameter and number of antennas within the network, diverse channel models and utilizing noisy channel estimates, and constrained ISAC problems.
 
	\bibliographystyle{IEEEtran}

\end{document}

%% file: input.tex
\usepackage{amsfonts}
\usepackage{times}
\usepackage{graphicx}
\usepackage{latexsym}
\usepackage{dsfont}
\usepackage{amssymb}
\usepackage{amsmath}
\usepackage{cite}
\usepackage{verbatim}
\usepackage{subfigure}

\newcommand{\figref}[1]{{Fig.}~\ref{#1}}

\def\bb0{{\mathbb{0}}}

\def\ba{{\mathbf{a}}}
\def\bb{{\mathbf{b}}}

\def\be{{\mathbf{e}}}

\def\bff{{\mathbf{f}}}

\def\bh{{\mathbf{h}}}

\def\bn{{\mathbf{n}}}

\def\bv{{\mathbf{v}}}

\def\bx{{\mathbf{x}}}
\def\by{{\mathbf{y}}}

\def\b0{{\mathbf{0}}}

\def\bE{{\mathbf{E}}}

\def\bG{{\mathbf{G}}}

\def\bI{{\mathbf{I}}}

\def\bW{{\mathbf{W}}}

\def\sf0{{\mathsf{0}}}









%% file: SPAWC_ISAC.bbl
\begin{thebibliography}{10}
		\providecommand{\url}[1]{#1}
		\csname url@samestyle\endcsname
		\providecommand{\newblock}{\relax}
		\providecommand{\bibinfo}[2]{#2}
		\providecommand{\BIBentrySTDinterwordspacing}{\spaceskip=0pt\relax}
		\providecommand{\BIBentryALTinterwordstretchfactor}{4}
		\providecommand{\BIBentryALTinterwordspacing}{\spaceskip=\fontdimen2\font plus
			\BIBentryALTinterwordstretchfactor\fontdimen3\font minus
			\fontdimen4\font\relax}
		\providecommand{\BIBforeignlanguage}[2]{{%
				\expandafter\ifx\csname l@#1\endcsname\relax
				\typeout{** WARNING: IEEEtran.bst: No hyphenation pattern has been}%
				\typeout{** loaded for the language `#1'. Using the pattern for}%
				\typeout{** the default language instead.}%
				\else
				\language=\csname l@#1\endcsname
				\fi
				#2}}
		\providecommand{\BIBdecl}{\relax}
		\BIBdecl
		
		\bibitem{liu2022integrated}
		F.~Liu, Y.~Cui, C.~Masouros, J.~Xu, T.~X. Han, Y.~C. Eldar, and S.~Buzzi,
		``Integrated sensing and communications: Towards dual-functional wireless
		networks for {6G} and beyond,'' \emph{IEEE J. Sel. Areas Commun.}, 2022.
		
		\bibitem{Demirhan_mgazine_radar}
		U.~Demirhan and A.~Alkhateeb, ``Integrated sensing and communication for {6G}:
		Ten key machine learning roles,'' \emph{IEEE Commun. Mag.}, 2023.
		
		\bibitem{ngo2017cell}
		H.~Q. Ngo, A.~Ashikhmin, H.~Yang, E.~G. Larsson, and T.~L. Marzetta,
		``Cell-free massive {MIMO} versus small cells,'' \emph{IEEE Trans. Wireless
			Commun.}, vol.~16, no.~3, pp. 1834--1850, 2017.
		
		\bibitem{demirhan2022enabling}
		U.~Demirhan and A.~Alkhateeb, ``Enabling cell-free massive {MIMO} systems with
		wireless millimeter wave fronthaul,'' \emph{IEEE Trans. Wireless Commun.},
		vol.~21, no.~11, pp. 9482--9496, 2022.
		
		\bibitem{behdad2022power}
		Z.~Behdad, {\"O}.~T. Demir, K.~W. Sung, E.~Bj{\"o}rnson, and C.~Cavdar, ``Power
		allocation for joint communication and sensing in cell-free massive {MIMO},''
		\emph{arXiv preprint arXiv:2209.01864}, 2022.
		
		\bibitem{demirhan2023cell}
		U.~Demirhan and A.~Alkhateeb, ``Cell-free {ISAC} {MIMO} systems: Joint sensing
		and communication beamforming,'' \emph{arXiv preprint arXiv:2301.11328},
		2023.
		
		\bibitem{zhou2020graph}
		J.~Zhou, G.~Cui, S.~Hu, Z.~Zhang, C.~Yang, Z.~Liu, L.~Wang, C.~Li, and M.~Sun,
		``Graph neural networks: A review of methods and applications,'' \emph{AI
			open}, vol.~1, pp. 57--81, 2020.
		
		\bibitem{liu2020joint}
		X.~Liu, T.~Huang, N.~Shlezinger, Y.~Liu, J.~Zhou, and Y.~C. Eldar, ``Joint
		transmit beamforming for multiuser {MIMO} communications and {MIMO} radar,''
		\emph{IEEE Trans. Signal Process.}, vol.~68, pp. 3929--3944, 2020.
		
		\bibitem{richards2010principles}
		M.~A. Richards, J.~Scheer, W.~A. Holm, and W.~L. Melvin, \emph{Principles of
			modern radar}.\hskip 1em plus 0.5em minus 0.4em\relax Citeseer, 2010, vol.~1.
		
		\bibitem{scarselli2008graph}
		F.~Scarselli, M.~Gori, A.~C. Tsoi, M.~Hagenbuchner, and G.~Monfardini, ``The
		graph neural network model,'' \emph{IEEE Trans. Neural Netw.}, vol.~20,
		no.~1, pp. 61--80, 2008.
		
		\bibitem{zhang2021scalable}
		X.~Zhang, H.~Zhao, J.~Xiong, X.~Liu, L.~Zhou, and J.~Wei, ``Scalable power
		control/beamforming in heterogeneous wireless networks with graph neural
		networks,'' in \emph{2021 IEEE GLOBECOM}, 2021, pp. 01--06.
		
		\bibitem{kim2022bipartite}
		J.~Kim, H.~Lee, S.-E. Hong, and S.-H. Park, ``A bipartite graph neural network
		approach for scalable beamforming optimization,'' \emph{IEEE Trans. Wireless
			Commun.}, vol.~22, no.~1, pp. 333--347, 2022.
		
		\bibitem{xu2023joint}
		C.~Xu, Y.~Jia, S.~He, Y.~Huang, and D.~Niyato, ``Joint user scheduling, base
		station clustering and beamforming design based on deep unfolding
		technique,'' \emph{IEEE Trans. Commun.}, 2023.
		
	\end{thebibliography}
